\documentclass[11pt]{article}
\usepackage[margin=1in]{geometry}
\usepackage[utf8]{inputenc}
\usepackage{amsmath,amssymb,amsfonts}
\usepackage{algorithmic}
\usepackage{graphicx}
\graphicspath{{experiments/}}
\usepackage{textcomp}
\usepackage{siunitx}
\usepackage{comment}
\usepackage{colortbl}
\usepackage{csquotes}
\usepackage{tabularx,booktabs}
\usepackage{float}
\usepackage[table]{xcolor}
\usepackage[hidelinks]{hyperref}
\usepackage{authblk}

\date{}

\newdimen\xfigwd \xfigwd=0pt

\usepackage{array}
\usepackage{adjustbox}

\definecolor{sc}{RGB}{255,235,59}    
\definecolor{ti}{RGB}{144,202,249}   
\definecolor{na}{RGB}{165,214,167}   
\definecolor{ph}{RGB}{244,143,177}   

\newcommand{\mnum}[2]{\cellcolor{#1}{\footnotesize #2}}

\begin{document}
\title{Practical Challenges in Executing Shor's Algorithm on Existing Quantum Platforms}
\author[1]{Paul Bagourd}
\author[1]{Julian Jang-Jaccard\thanks{Corresponding author: julian.jang-jaccard@ar.admin.ch}}
\author[1]{Vincent Lenders}
\author[1]{Alain Mermoud}
\author[2]{Torsten Hoefler}
\author[3]{Cornelius Hempel}
\affil[1]{armasuisse Science and Technology, Cyber-Defence Campus, Thun, Switzerland}
\affil[2]{ETH Zurich, Department of Computer Science, Zurich, Switzerland}
\affil[3]{Paul Scherrer Institute, ETH Zurich--PSI Quantum Computing Hub, Villigen, Switzerland}

\maketitle

\begin{abstract}
Quantum computers pose a fundamental threat to widely deployed public-key cryptosystems, such as RSA and ECC, by enabling efficient integer factorization using Shor's algorithm.
Theoretical resource estimates suggest that 2048-bit RSA keys could be broken using Shor’s algorithm with fewer than a million noisy qubits. Although such machines do not yet exist, the availability of smaller, cloud-accessible quantum processors and open-source implementations of Shor’s algorithm raises the question of what key sizes can realistically be factored with today's platforms. 
In this work, we experimentally investigate Shor's algorithm on several cloud-based quantum computers using publicly available implementations. Our results reveal a substantial gap between the capabilities of current quantum hardware and the requirements for factoring cryptographically relevant integers. 
In particular, we observe that circuit constructions still need to be highly specific for each modulus, and machine fidelities are unstable with high and fluctuating error rates.
Although our findings indicate that the present state of quantum computing poses no immediate or near-term threat to modern cryptography, a combination of breakthroughs in hardware and algorithmic advances could rapidly change this picture. Continuous monitoring and proactive preparation are therefore essential to stay ahead of emerging quantum threats.

\end{abstract}

\noindent\textbf{Keywords:} cryptography, prime factorization, public key, quantum computing, Shor's algorithm

\section{Introduction}
\label{sec:introduction}
Quantum computers have the potential to transform computing by efficiently solving certain problems that are intractable for classical machines. Shor's algorithm~\cite{b1,b2} is particularly notable for its ability to factor large integers in polynomial time, outperforming the best-known classical methods. At the core of Shor's algorithm is the Quantum Fourier Transform (QFT), which enables key subroutines to be implemented exponentially faster than their classical counterparts. This capability threatens widely used public-key cryptographic protocols, such as RSA and also ECC~\cite{b29}, whose security is based on the presumed hardness of integer factorization or related number-theoretic problems.
While theoretical advancements have clarified how quantum computers could eventually compromise such systems, the practical implementation of Shor's algorithm on existing quantum hardware remains extremely challenging — even for integers far smaller than those used in deployed cryptographic schemes. In the current \emph{noisy intermediate-scale quantum} (NISQ) \cite{preskill_quantum_2018} era, quantum devices have limited qubit counts, significant noise, and short coherence times. Understanding what they can and cannot realistically achieve is crucial for assessing the urgency of quantum-safe cryptographic migration.

This paper makes three main contributions. 
First, we review the leading quantum computing approaches that have reached comparable technical readiness levels. We group them into synthetic and natural qubits and discuss their strengths, weaknesses, and key technical barriers.

Second, we review the state of experimental implementations of Shor's algorithm across various quantum computing technologies attempted to date, discussing the significance and limitations of each effort.

Third, we report on our own experiments in performing order finding -- the core quantum subroutine of Shor’s algorithm -- on a publicly available superconducting quantum processor. Using carefully engineered circuits and statistical analysis of Quantum Phase Estimation (QPE) histograms, we characterize the observable quantum signal and identify the point where noise overwhelms the algorithmic structure. We complement these results with an investigation of the challenges encountered when attempting to run comparable circuits on non-superconducting platforms via cloud services.

Overall, our findings suggest that, despite substantial progress in both algorithms and hardware, practical quantum attacks on standard cryptographic key sizes remain out of reach on currently accessible quantum platforms. However, the rapid pace of development, coupled with ambitious industrial roadmaps, underscores the need for ongoing evidence-based assessments of quantum capabilities.

\section{Understanding Different Quantum Computers}
Quantum computing is built on qubits as the fundamental units of quantum information. Broadly, qubit implementations can be categorized into two families: synthetic qubits, which are artificially engineered entities designed to effectively realize two-level quantum behavior, and natural qubits, which directly use the intrinsic quantum states of atoms, ions, or photons. 
Table \ref{tab:qubit-tech} provides an overview of various key qubit technologies, representative commercial providers, and high-level performance characteristics.

\begin{table*}[ht]
\caption{Quantum computing with different qubit technologies and their characteristics.}
\label{tab:qubit-tech}
\footnotesize
\setlength{\tabcolsep}{5pt}
\begin{tabularx}{\textwidth}{
  >{\raggedright\arraybackslash}p{0.19\textwidth}
  >{\raggedright\arraybackslash}X
  >{\raggedright\arraybackslash}X
  >{\raggedright\arraybackslash}X}
\toprule
 & \textbf{Synthetic Qubits} & \multicolumn{2}{c}{\textbf{Natural Qubits}} \\
\cmidrule(lr){3-4}
 &  & \textbf{Atom-based} & \textbf{Light-based} \\
\midrule
\textbf{Type} &
Superconducting circuits*, Quantum dots, Topological qubits, NV centers in diamonds &
Trapped-Ion*, Neutral Atom* &
Photon \\
\midrule\textbf{Examples of Commercial Providers} &
IBM, Google, Rigetti, IQM, Microsoft, Intel, Diraq, Quantum Brilliance &
IonQ, Quantinuum, AQT, QuEra, Pasqal, planqc, Atom Computing, Infleqtion &
Xanadu, PsiQuantum, Quandela \\
\midrule\textbf{Overall Processing Speed} & Fast & Slow & Fast \\
\midrule\textbf{Qubit Connectivity} & Low & High & Low \\
\midrule\textbf{Coherence Times} & Low & High & High \\
\midrule
\textbf{Cooling Requirement} &
Extreme cryogenics ($\sim$10 mK) &
Room temperature but often standard cryogenics ($\sim$4 K) &
Room temperature with detectors requiring standard cryogenics (4 K) \\
\midrule
\textbf{Main Advantages} &
Circuits can be designed with the desired topology; highly scalable in some approaches &
High accuracy and stability; qubits identical by nature; very long coherence times &
Energy efficient; builds on some telecom-industry matured photonic technologies \\
\midrule
\textbf{Main Challenges} &
Qubit uniformity; limited cooling power and wiring; high sensitivity to noise; fast classical-electronics requirements &
Complex laser-based control that is hard to scale &
Photon loss and detection; entanglement preparation; often non-deterministic operation slowing clock speeds \\
\bottomrule
\end{tabularx}

\vspace{2pt}\emph{* indicates the highest TRL modalities at the time of writing}
\end{table*}

\subsection{Quantum Computers with Synthetic Qubits}
Superconducting qubits, widely used by companies such as IBM, Google, and Rigetti, encode information in oscillating electrical currents in superconducting circuits cooled to millikelvin temperatures. They offer fast gate operations and benefit from mature microwave control and fabrication ecosystems. However, they suffer from relatively short coherence times, variability across the chip, and significant sensitivity to noise, all of which increase the burden on error mitigation and QEC.

Quantum dots, which confine electrons in semiconductor nanostructures, are attractive due to their compatiblity with existing semiconductor technology and their potential for high-density integration. Their main challenges include short coherence times, tight fabrication tolerances, and complex control requirements. The availability of ultra-pure silicon-28 as raw material can provide further limitations. 

Topological qubits, based on exotic quasiparticles such as Majorana fermions, are predicted to exhibit intrinsic robustness against certain types of local noise, thereby reducing error correction overheads in principle. Yet, their experimental realization remains difficult and has not been conclusively demonstrated.

NV centers in diamond use electronic or nuclear spin associated with a lattice defect to encode qubit states. They offer long coherence times and can operate at room-temperature, but scaling them to large, controllable registers is difficult, due to the difficulty of precise and reliable manufacturing.

Overall, synthetic qubits benefit from strong synergies with microelectronics and integrated circuits, but progress towards large-scale, high-fidelity systems is hindered by noise, parameter dispersion, and packaging constraints that become more severe as system size grows.

\subsection{Quantum Computers with Natural Qubits}
Natural qubit platforms can be further divided into atom-based and light-based approaches.

Trapped-ion systems encode qubits in the internal electronic states of ions confined in electromagnetic traps under ultra-high vacuum. They are renowned for their long coherence times and record-low gate error rates, and have been the first platform to demonstrate deterministic quantum gate operations already in 1995. Today, companies such as IonQ and Quantinuum have demonstrated programmable systems with tens of high-fidelity qubits. However, gate operations are relatively slow, and scaling brings challenges due to the complexity trap architectures and laser delivery.

Neutral atom platforms trap individual atoms in optical potentials generated by tightly focused laser beams or optical latices. These systems promise high scalability, as atoms, like ions, are identical and can be arranged in dense arrays. At the same time, comparably weak trapping restricts interactions to nearest neighbors, requiring time-consuming qubit rearrangements, and makes these systems highly sensitive to losses from collisions with background gas. Precise control over large arrays remains a challenge because of the need for very high laser powers for optical trapping, combined with complex optical setups for rearrangements.

Photonic quantum computing systems use the quantum properties of light to encode qubits. Here, typically the degrees of freedom of single photons, such as polarization are exploited. Photons are naturally well-suited for quantum communication because of their weak coupling to the environment and their ability to travel over long distances. Companies such as  Xanadu and PsiQuantum pursue architectures building on technology adapted from classical telecommunications. Yet, realizing scalable photonic processors is difficult as photon-loss and the difficulty of generating entangled photon pairs (a probabilistic process) reduce effective clock speeds and high measurement efficiency requires low temperature cryogenics for the photon detectors.

Together, these approaches highlight a diverse technology landscape, in which no single platform currently dominates on all relevant metrics (qubit count, fidelity, connectivity, speed, and compatibility with QEC). This diversity is directly relevant to the practical implementation of Shor’s algorithm.

\section{Review on Shor's Implementations}
\subsection{Shor's Algorithm}
As discussed in \cite{b18}, quantum computers excel at solving problems with modest classical input/output data but high computational complexity. Shor’s algorithm \cite{b1,b2} is a canonical example: it factors an $n$-bit integer $N$ in time polynomial in $n$, in stark contrast to the sub-exponential but super-polynomial complexity of the best-known classical factoring algorithms.

At a high level, Shor’s algorithm consists of three main steps:
\begin{enumerate}
  \item \textbf{Choose a random base (classical). \label{step1} } Pick an integer \(a\) with \(1<a<N\) and \(\gcd(a,N)=1\). Let \(r\) be the smallest positive integer such that
  \[
    a^{r}\equiv 1 \pmod{N},
  \]
  i.e., the \emph{period} of the sequence \(a^{x}\bmod N\).

  \item \textbf{Find the period (quantum).\label{step2} } Use a QFT–based QPE routine to infer \(r\) from the periodic structure of \(a^{x}\bmod N\).Classically, order finding in $\mathbb{Z}_N^{\times}$ is not known to admit a polynomial-time algorithm and is believed to require sub-exponential time in the bit-length of $N$. In contrast, a fault-tolerant quantum computer can solve this problem in polynomial time.

  \item \textbf{Compute the factors (classical).\label{step3} } If \(r\) is even and \(a^{r/2}\not\equiv -1 \pmod N\), then
  \[
    \gcd\!\big(a^{r/2}-1,\,N\big)\quad\text{and}\quad \gcd\!\big(a^{r/2}+1,\,N\big)
  \]
  yield nontrivial factors of \(N\). Otherwise, choose a new \(a\) and repeat.
\end{enumerate}

\subsection{Shor's Implementation: State-of-the-Art}
In 2001, a group of IBM researchers presented the earliest experimental demonstration of Shor's algorithm, successfully factoring $15= 3 \times 5$ using a seven-qubit liquid-state \footnote{Uses nuclear spins in molecules dissolved in a liquid as qubits.} nuclear magnetic resonance (NMR) quantum computer~\cite{b3}. While liquid-state NMR quantum computers are not considered a scalable route for solving larger problems, the experiment provided an important proof of principle, demonstrating the algorithm's feasibility.

Subsequent experiments implemented compiled versions of Shor’s algorithm using photonic qubits, again targeting the factorization of 15~\cite{b4,b5}. These demonstrations focused on generating and characterizing multi-qubit entanglement in circuits derived from Shor’s algorithm, but relied on heavy compilation that exploits prior knowledge of the factors.

In 2012, factorization of 15 was successfully demonstrated on a superconducting quantum processor~\cite{b6}. This milestone marked a significant advancement, showcasing the feasibility of implementing Shor's algorithm on a quantum platform aligned with long-term scalability goals. In the same year, a photonic implmentation successfully factored 21~\cite{b7}.

In 2016, factorization of 15 was achieved using a trapped-ion implementation which used a qubit recycling technique~\cite{b8}. 
Unlike previous compiled approaches, this demonstration did not rely on prior knowledge of the prime factors, and the design was argued to be scalable in principle, albeit still limited to a four-bit composite.

More recent work was carried out on larger superconducting devices.
In 2019, an attempt was made to factor 35 on the IBM Q System One~\cite{b9}. However, the algorithm achieved only a 14\% success rate due to accumulated noise, illustrating the fragility of deep circuits on current hardware. The latest IBM-based demonstation adopts a hybrid quantum-classical technique to variationally factor 253 (an 8 bit composite)~\cite{b30}. While such methods are promising, they change the algorithmic structure and require careful interpretation when extrapolating to cryptographic scales.

A number of works claim factorizations of larger integers using Shor-like circuits. As emphasized in the aptly-titled manuscript \enquote{Oversimplifying Quantum Factoring}~\cite{b12}, many of these experiments rely on heavily compiled circuits that incorporate knowledge of the answer, drastically lowering resource requirements, and sometimes reducing the problem to something closer to classical coin flipping than genuine quantum factoring. This underscores the importance of benchmarking implementations that preserve the true complexity of order-finding.

In parallel, classical simulations of Shor's algorithm on large-scale high-performance computing systems have advanced significantly. Recent work~\cite{b19} reached 39 bits RSA key length on GPU clusters, validating the algorithmic structure and resource trends in a controlled environment. Nevertheless, the current (public) record is still held by classical algorithms such as the \enquote{Generalized Number Field-Sieve} (GNFS) which has reached 829 bits of key length~\cite{b25}.

On the algorithmic side, the past two years have seen the first notable developments in the 30 years since Shor's discovery in 1994. 
 Recent work has lowered qubit counts~\cite{b20,b21,b26} and improved asymptotic runtime~\cite{b22}, particularly by leveraging advances in quantum error correction and more efficient order-finding techniques.
  An important point to observe is the growing efficiency of incorporating progress in quantum error correction, which can be considered as the software layer between hardware (noisy physical qubits) and application (factoring). 
 For example, updated resource estimates for factoring RSA-2048 have improved from around 20 million qubits and 8 hours of execution time~\cite{b17} to fewer than one million qubits and roughly 5 days~\cite{b27}, under realistic assumptions about error rates and QEC overhead. 
  Architectures that combine novel qubit designs with multimode memories could, at least in principle, reduce the required qubit count by another two orders of magnitude, down to approximately 13\,436 qubits~\cite{b28}.
These developments emphasize that both hardware and algorithms are progressing and that resource estimates continue to evolve.

\section{Shor's Implementation on Existing Quantum Platforms}

\subsection{Setup and Notation}
Shor's expected runtime factors into two quantities: the cost of a single order-finding run and the number of independent runs with random bases $a$ required until a non-trivial factor appears in the post-classical step. Let $C_{\mathrm{run}}(N)$ be the runtime of one order-finding attempt and let $p_{\mathrm{succ}}(N)$ be the per-run success probability. The expected number of repetitions $R$ is $\mathbb{E}[R]=1/p_{\mathrm{succ}}(N)$, so the total expected runtime is:
\begin{equation}
\mathbb{E}[\mathrm{time}] \;=\; \frac{C_{\mathrm{run}}(N)}{p_{\mathrm{succ}}(N)}. \label{eq:expected_time}
\end{equation}

With fast modular arithmetic, $C_{\mathrm{run}}(N)=\tilde O((\log N)^2)$ (i.e., polylogarithmic in $N$ and polynomial in the bit-length of $N$). Thus, certifying that Shor’s algorithm runs in polylogarithmic expected time reduces to experimentally lower bounding $p_{\mathrm{succ}}(N)$.

Let $N$ be the composite to factor, $a$ a random coprime modulo $N$, and $r$ the order of $a \bmod N$. The phase register in QPE has $t$ qubits, so its grid has size $L=2^t$ and the measured outcome is $y\in\{0,\dots,L-1\}$. In the ideal case, QPE produces peaks near the rationals $s/r$ for $s=0,\dots,r-1$. We choose, denoting $n=\lceil\log_2 N \rceil $ the number of bits,
\begin{equation}
t \;\ge\; 2n\label{eq:t_choice}
\end{equation}
so that continued fractions (CF) can correctly recover $r$ from sufficiently precise samples, even in the worst-case scenario, without relying on prior knowledge about $N$.\\
However, with modern post-processing (Ekerå ~\cite{b31,b32}), the required phase precision can be cut to roughly $t \approx n + O(\log n)$ while maintaining high single-run success via limited classical searches, thereby reducing depth and repetitions. Thus, we allow ourselves to use a lower $t=10<12$ for the case $N=35$ while still considering the framework scalable \footnote{Further details on the experimental certification method and refinements regarding scalability in the context of Shor's algorithm are provided in the Appendix~\ref{appendix}.}.

\subsection{Public Quantum Platforms}

For our study, we used the IBM Quantum Platform\footnote{https://quantum.ibm.com/} which provides public, cloud-based access to a family of superconducting quantum processors.\footnote{https://docs.quantum.ibm.com/}
All circuits were created in Jupyter notebooks using Qiskit, transpiled to match the topology and native gate set of the chosen backend, and executed using the IBM Quantum Runtime (Sampler V2). We did not implement manual error correction; instead, we relied on the platform’s built-in error mitigation and calibration procedures.


\begin{table}[h]
\centering
    \centering
    \caption{Specifications of the \texttt{ibm\_torino} Quantum Processor}
    \label{table:ibm_torino_specs}
    \begin{tabular}{@{} l l @{}}
        \toprule
        \textbf{Specification} & \textbf{Value} \\
        \midrule
        Machine Name & \texttt{ibm\_torino} \\
        Total Qubits & 133 \\
        Processor (Revision) & Heron (r1) \\
        Public Debut Date & 2023-12-04 \\
        Currently Available & Yes \\
        \midrule 
        $T_1$ ($\mu$s) & \num{188.48} \\
        $T_2$ ($\mu$s) & \num{140.66} \\
        \midrule 
        1Q Error (SX Gate) & \num{2.876e-04} \\
        2Q Error (CX Gate) & \num{2.733e-03} \\
        Readout (RO) Error & \num{2.917e-02} \\
        \bottomrule
    \end{tabular}
\end{table}

Table~\ref{table:ibm_torino_specs} summarizes the characteristics of \texttt{ibm\_torino}, the backend used in our experiments. Two coherence-time metrics are listed. $T_1$ (energy relaxation time) measures how long a qubit prepared in the excited state ($\vert{1}\rangle$) remains there before decaying to the ground state ($\vert{0}\rangle$). Longer $T_1$ values indicate reduced energy relaxation and better stability. 
$T_2$ (dephasing time) quantifies how long a qubit maintains phase coherence in a superposition state; it captures the rate at which quantum information is lost due to environmental noise and control imperfections. While $T_2$ is upper-bounded by $2\,T_1$, it is typically significantly shorter in practice due to imperfect control and environmental noise.

The three error-rate metrics quantify the gate and measurement performance of the machine. The \emph{1Q Error (SX Gate)} is the error probability of the single-qubit $X_{\pi/2}$ (``SX'') rotation (typically derived from randomized benchmarking). The \emph{2Q Error (CX Gate)} is the error probability of the platform's native two-qubit entangling gate (an echoed cross-resonance gate implementing a CNOT-like primitive). The \emph{Readout Error (RO)} quantifies the probability of misclassifying the final measurement outcome ($\vert{0}\rangle\leftrightarrow\vert{1}]\rangle$).
These values reflect native, physical error rates without QEC; lower values permit deeper and wider circuits before noise dominates.


\paragraph*{Code}
Our implementation is based on standard constructions for reversible arithmetic. We employ a Cuccaro-style ripple-carry adder~\cite{Cuccaro2004} and a two's-complement overflow/comparator method for modular addition~\cite{Oumarou2022}, avoiding QFT-space adders~\cite{Draper2000}. Controlled powers of $a$ are implemented via precomputed exponents $a^{2^k}\bmod N$ within a textbook QPE order-finding scaffold~\cite{Kitaev1997,Cleve1998} and we target Qiskit Runtime Sampler V2~\cite{QiskitRuntime}.

\paragraph*{Implementation Approaches}
We derived an implementation of order-finding via Quantum Phase Estimation (QPE) that avoids the ad-hoc modular exponentiation tailored to a specific $N$.
Instead of hard-coding multiply-by-a mod N for a fixed N, we programmatically assemble a permutation operator on the $2^n$-dimensional space that maps $x \to (a\cdot x) \bmod N$ for $x < N$ and acts as identity otherwise, preserving unitarity. Controlled powers $U^{2^k}$ are constructed from $(N, a, t)$, and we implement a manual inverse QFT followed by bit-order swaps, ensuring that the measured bitstrings are consistently interpreted across Qiskit versions.

We implement the \emph{parallel} (non-iterative) phase-estimation variant of Shor’s order-finding rather than Kitaev’s iterative version \cite{Kitaev1997}. Both are algorithmically equivalent: the dominant cost is the controlled modular exponentiation \(U_a\), which scales as \(\tilde{O}(n^3)\) gates for \(n=\lceil \log_2 N\rceil\) with standard reversible adders and modular reduction, while the inverse QFT contributes \(O(t^2)\) gates for \(t\) phase bits. The choice is therefore an engineering trade-off. Parallel QPE uses \(t\) control qubits and a single inverse QFT, yielding a full \(t\)-bit outcome per shot. Iterative QPE reuses one control qubit over \(t\) rounds with mid-circuit measurement and classical feed-forward, thereby reducing qubit count but increasing depth adding additional SPAM/latency errors. In our experiments, \(t\approx 2\lceil \log_2 N\rceil\) is modest, so the extra control qubits are acceptable and let us (i) avoid real-time feedback, (ii) exploit vectorized sampling, and (iii) apply a uniform multi-peak histogram analysis pipeline based on acceptance windows and binomial tests. In a fault-tolerant setting, either variant remains viable; our choice is driven by hardware constraints, not complexity. 

The same circuit runs unchanged on Aer or IBM Runtime (SamplerV2). Signal quality is quantified by windowed hit-rate versus the uniform baseline.  For larger $N$ $(N>30)$, the convenient and generic dense permutation can be replaced by structured modular-arithmetic circuits without changing the interface, preserving the generic pipeline while improving scalability.

\subsection{Results (N=15, 21, 35)}
The following histograms illustrate the measured results after performing quantum phase estimation repeatedly. In all figures, the x-axis represents the binary bitstrings of the first eight qubits used in the measured outcome, corresponding to the quantum states after executing the quantum circuit. The y-axis represents how often each measurement outcome was observed.

\begin{figure}[H]
  \centering
  \includegraphics[width=\columnwidth]{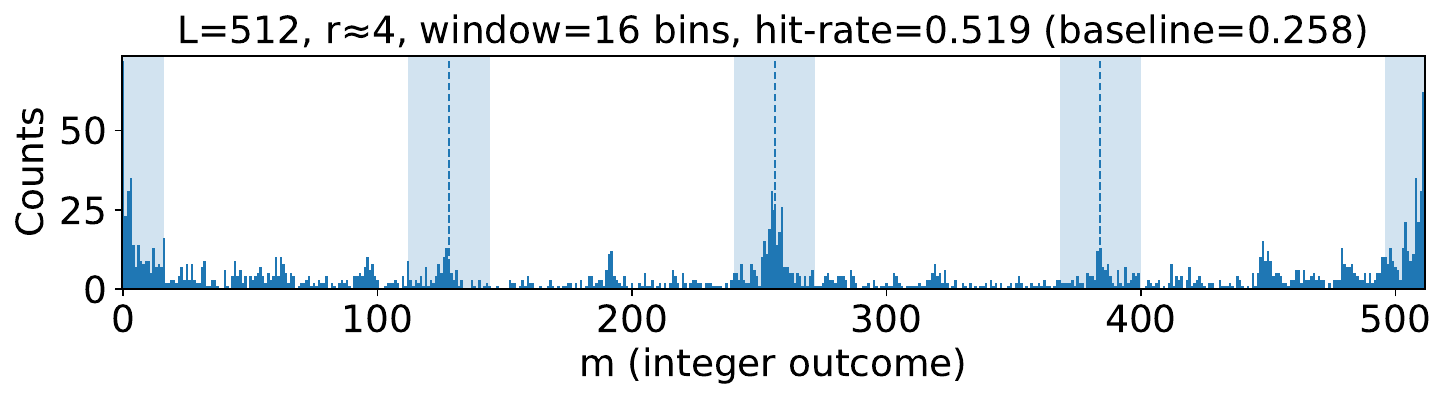}
  \caption{Measured QPE histogram for $N=15$ (simulation).}
  \label{fig:N15simul}
\end{figure}

We first validated our implementation  on an ideal quantum simulator running on a classical workstation. In this noiseless setting, every gate is executed perfectly with no decoherence, crosstalk, readout errors, or noise model, so the only randomness is from finite shots and the intrinsic QPE Fourier spread. As a result, one still see some counts outside the acceptance windows: QPE produces tails around each peak, and with finite shots those tails appear as small bars in the out‑of‑window region even in the ideal simulator. As shots tend to infinity on the ideal simulator: the empirical histogram converges to the exact QPE distribution. The peak locations stay the same, the peak heights approach their true probabilities, and the statistical noise vanishes. For instance, the simulation for $N=15$ with parameters $t=9$, $a=7$, $shots=2048$, $r=4$ gives the following histogram \ref{fig:N15simul}.  The baseline is the fraction of bins covered by the acceptance windows (so a uniformly random outcome would land there with that probability).  The hit-rate obtained is twice the uniform baseline rate, suggesting a strong simulated quantum signal.

We then executed the same circuits on  \textit{ibm\_torino}, after transpilation to the device topology. 
For each modulus $N$, we focus on detecting a statistically significant excess of probability mass in the theoretically predicted windows around $sL/r$ relative to the uniform baseline. The formulas of the uniform baseline and the empirical probability used for the experiments, rigorously derived in the Appendix \ref{appendix}, are the following:
\[b=\frac{\vert acceptance\_bin\vert \times (2\omega_0+1)}{L} \text{ and }\hat{p}=\frac{hits}{shots}\] 
We consider three composite numbers: $N=15$, $N=21$, and $N=35$.  In the Appendix subsection \ref{effect_decoherence}, we further discuss the shape of the empirical histograms compared to the ideal noiseless expected distributions.

\begin{figure}[H]
  \centering
  \includegraphics[width=\columnwidth]{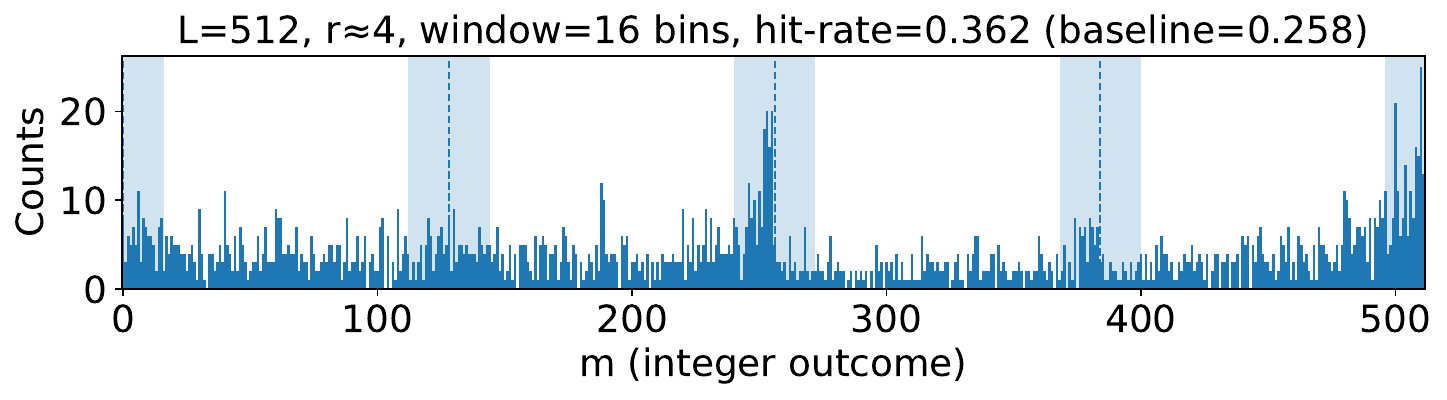}
  \caption{Measured QPE histogram for $N=15$.}
  \label{fig:N15}
\end{figure}

\paragraph*{N=15 $\rightarrow$ PASS}
With parameters $t=9$, $a=7$, $shots=2048$, $r=4$, we obtain the histogram \ref{fig:N15}.

With inclusive windows, each peak spans $(2w_0+1)=33$ bins, so the total number of accepted bins is $4\times 33=132$, yielding a uniform baseline $b=\frac{132}{512}=0.258$. From $n=2048$ shots, we observe $k=741$ hits inside the acceptance windows, corresponding to $\hat{p}=0.362$ and  an excess $\hat{p}-b=0.104$. A one-sided binomial test for $H_0:\,p\le b$ vs.\ $H_1:\,p>b$ returns $p_{\mathrm{val}}=3.50\times 10^{-27}$, so at the significance level $\alpha=0.01$ we reject $H_0$ and conclude that the data exhibits a statistically significant quantum signal ($N=15\rightarrow$ PASS).

\begin{figure}[H]
  \centering
  \includegraphics[width=\columnwidth]{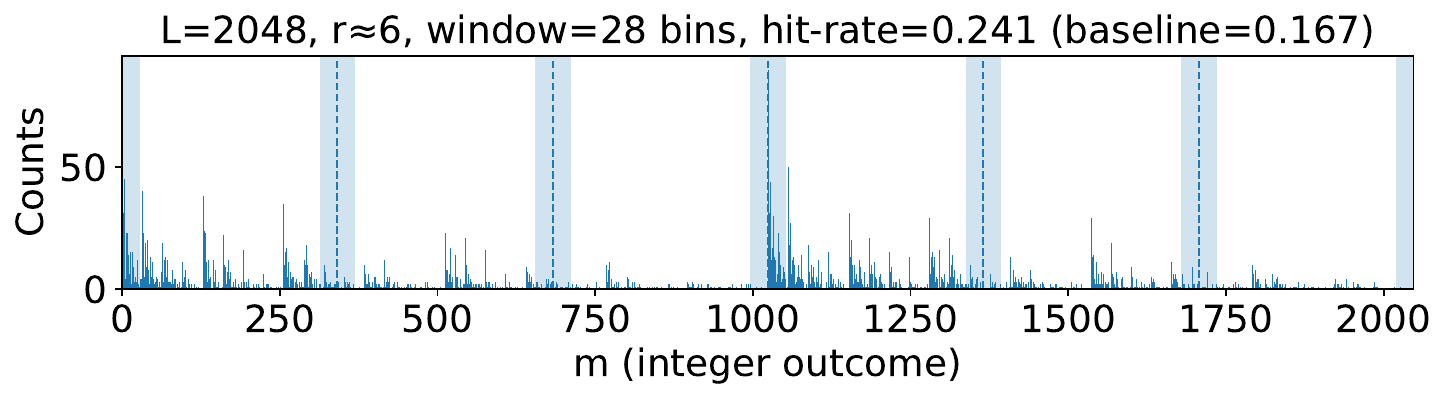}
  \caption{Measured QPE histogram for $N=21$.}
  \label{fig:N21}
\end{figure}

\paragraph*{N=21 $\rightarrow$ PASS}
With parameters $t=11$, $a=2$, $shots=4096$, $r=6$, we obtain the histogram \ref{fig:N21}.
Each peak window spans $(2w_0+1)=57$ bins, so the total number of accepted bins is $6\times 57=342$, giving a baseline $b=\frac{342}{2048}=0.167$. We observe $k=988$ hits inside the windows such that $\hat{p}=0.241$ and excess $\hat{p}-b=0.074$. The one-sided binomial test for $H_0:\,p\le b$ vs.\ $H_1:\,p>b$ yields $p_{val}=2.45\times 10^{-37}$, so again at $\alpha=0.01$ we reject $H_0$ and detect a strong quantum signal ($N=21\rightarrow$ PASS).

\begin{figure}[H]
  \centering
  \includegraphics[width=\columnwidth]{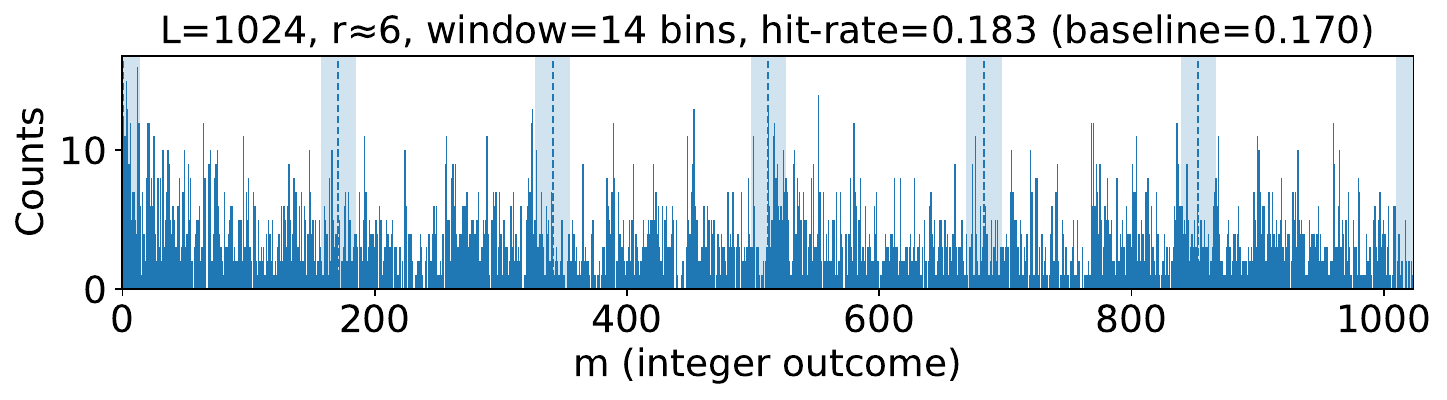}
  \caption{Measured QPE histogram for $N=35$ $(a=4)$.}
  \label{fig:N35_hard}
\end{figure}

\paragraph*{N=35 (two experiments)}
In the first experiment, we set $t=10$, $a=4$, $shots=4096$, $r=6$ to obtain histogram \ref{fig:N35_hard}.

Each peak spans $(2w_0+1)=29$ bins, so there are $6\times 29=174$ accepted bins, giving $b=\frac{174}{1024}=0.170$. From $n=4096$ shots, we record $k=751$ hits inside the windows, yielding $\hat{p}=0.183$ and  $\hat{p}-b=0.013$. The corresponding one-sided binomial test gives $p_{val}=1.17\times 10^{-2}$, so at $\alpha=0.01$ we do not reject $H_0$, finding only marginal evidence for a quantum signal (FAIL).\\

\begin{figure}[H]
  \centering
  \includegraphics[width=\columnwidth]{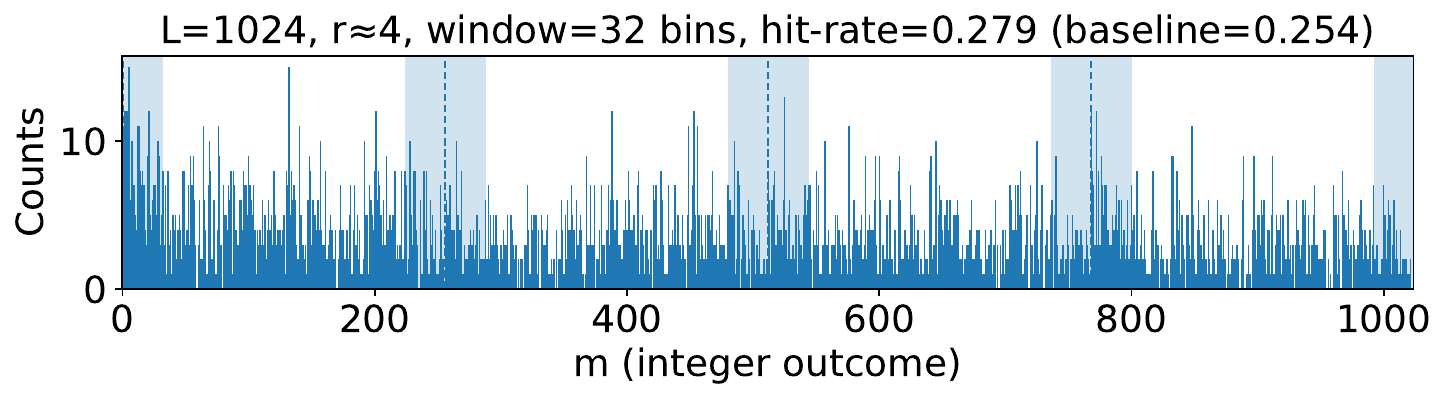}
  \caption{Measured QPE histogram for $N=35$ $(a=8)$.}
  \label{fig:N35_easier}
\end{figure}

In the second experiment, we use a different base $a=8$ with parameters $t=10$, $shots=4096$, $r=4$ to obtain histogram \ref{fig:N35_easier}.

Here, each peak spans $(2w_0+1)=65$ bins, so $4\times 65=260$ bins are accepted, giving $b=\frac{260}{1024}=0.253$. With $k=1144$ hits inside the windows, we obtain $\hat{p}=0.279$ and excess $\hat{p}-b=0.0253$. The binomial test yields $p_{val}=1.17\times 10^{-4}$, so we reject $H_0$ at $\alpha=0.01$ and again detect a quantum signal, albeit much weaker (PASS).\newline

The significant change in performance across the two experiments is explained as follows. The choice of base $a$ (coprime to $N$) fixes the order $r$ and thus the circuit depth/structure and peak geometry. “Friendly” $a$ can yield smaller $r$, shallower controlled-$U^{2^k}$ ladders, fewer two-qubit errors, and peaks that align well with $L=2^t$ (narrower, higher peaks, larger excess $\hat p-b$). “Unfriendly” $a$ can give larger $r$ or poor alignment with $L$, leading to deeper circuits, more noise accumulation and broader peaks, shrinking the excess $(\hat p-b)$. The cases $N=15$ and $N=21$ use deliberately unfriendly bases $a$ to stress-test robustness in a worst-case scenario. In contrast, $N=35$ exhibits mixed outcomes: unfriendly base $a=4$ give borderline detection, while more friendly base $a=8$ passes but with a weak signal. Rigorous studies should disclose how $a$ is chosen, and, ideally, include worst-case or at least non–cherry-picked bases.

\paragraph*{Conclusions of the experiments}
For $N=15$ and $N=21$, the observed hit rates substantially exceed the baseline, providing clear statistical evidence of QPE peaks at the expected locations. For $N=35$, the excess is modest, yielding only marginal evidence of a quantum signal, and depends sensitively on the choice of base $a$. This indicates that we are near the practical limit of what the device can support with our circuit design and shot budget, aligning with expectations for NISQ hardware, where noise and limited coherence quickly degrade performance as circuit-width ($\propto \#\mathrm{qubits}$) and -depth ($\propto \#\mathrm{gates}$) grow.

\subsection*{Post-processing part (classical)}
Once a period $r$ is successfully recovered from the QPE output via continued fractions, the classical step of Shor’s algorithm computes $\gcd(a^{r/2}\pm 1,N)$ and checks whether this yields a non-trivial factor. As discussed earlier, this post-processing is not the algorithmic bottleneck and runs in polynomial time on a classical computer. Accordingly, we do not elaborate further on this component.

\subsection*{Other Experimental Attempts}
We also attempted to implement Shor’s algorithm on quantum devices beyond superconducting-based qubit systems. Amazon Braket \footnote{https://aws.amazon.com/braket/}, a cloud-based quantum service from AWS, provides access to quantum hardware from multiple vendors, including IonQ (trapped ions), IQM and Rigetti (superconducting circuits), and QuEra (neutral atoms). 
We used Braket’s Qiskit-compatible interface to avoid major code rewrites. While our code ran successfully on Braket’s simulator, the platform failed to transpile the circuits to match the architecture of any available quantum hardware, possibly due to limits on depths or native gate sets.

A similar attempt was made with Microsoft Azure Quantum\footnote{https://quantum.microsoft.com/}, which provides access to IonQ, Quantinuum (trapped ions), and Rigetti (superconducting circuits). Using the Q\# interface, we ran our Qiskit-based Shor’s algorithm, but as with Braket, circuit transpilation to physical hardware did not succeed, preventing us from evaluating the algorithm on non-superconducting circuit platforms.

These attempts highlight the difficulty of porting non-trivial quantum algorithms across heterogeneous hardware platforms using generic transpilation pipelines. Publicly available code that runs successfully on IBM hardware does not automatically translate to devices with different qubit topologies, native gates, and other constraints. In practice, achieving reliable transpilation and execution on non-IBM platforms appears to require hardware-specific circuit design and optimization, which lies beyond the scope of this work. Moreover, while Qiskit includes a built-in pedagogical implementation of Shor’s algorithm, it is not designed to scale to realistically large integers or to serve as a drop-in attack tool on current NISQ devices.

Taken together, these observations suggest that there is currently no practical, platform-agnostic \enquote{Shor package} capable of factoring even moderately larger integers beyond standard textbook examples such as 15 or 21. The combination of algorithmic complexity, qubit noise, limited connectivity, and immature transpilation tooling remains a major hurdle.

\section{Limitations and Challenges}

\subsection*{Limitations with Circuit Design for Each N}
Our implementation is \emph{semi-generic}: a single generator produces the QPE scaffold and reversible arithmetic for any target bit-width $n$ and base $a$, but the modulus $N$ and derived constants (e.g., \(\bar N=2^{n}-N\) and \(k_i=c\,2^i \bmod N\)) are embedded directly into the gate patterns. As a result, each pair $(N,a)$ yields a distinct circuit that must be transpiled. 

To achieve \emph{genericity at fixed width}, the arithmetic would need to treat $N$ and related parameters as \emph{data} rather than as compile-time \emph{structure}. This would entail allocating an $n$-qubit modulus register \(\mathrm{NREG}\) (and addend/multiplier registers as needed), replace constant-injection adders with register–register Cuccaro adders plus a reversible subtract-and-borrow comparator against \(\mathrm{NREG}\), and implement modular multiplication as a variable shift-and-add with reductions modulo \(\mathrm{NREG}\) (optionally via parameterized Fourier-space adders)~\cite{b11}. Consequently, a comprehensive benchmark should employ $N$-agnostic arithmetic register–register add/compare/multiply mod $N$ or parameterized Fourier-space rotations so that a single width-fixed topology works for any $N<2^n$. Per-$N$ specialized circuits primarily demonstrate the behavior of bespoke instances and do not fully probe a device’s readiness for the generic operations required at cryptographically relevant key sizes.

\subsection*{Limitations with Machine Fidelity}
Quantum machine fidelity captures how accurately a device implements its nominal operations. Other factors such as qubit connectivity, coherence times, and readout error rates provide further metrics to estimate whether reliable results are produced when running Shor's algorithm on real hardware. While we had access to calibration data for \texttt{ibm\_torino} (Table~\ref{table:ibm_torino_specs}), these values represent median performance and can vary substantially over time. In practice, we observed non-negligible day-to-day fluctuations that occasionally degraded circuit behavior to the point of rendering runs effectively unusable.
As a result, some experiments had to be discarded as outliers corresponding to unfavorably calibrated device states. Below are presented the statistics of some error metrics over the months preceding and succeeding the experiments, for qubit \textit{q\_0}. A more systematic characterization of worst-case behavior-for instance, via repeated sampling over many calibration cycles-would be necessary to rigorously assess reliability guarantees. However, such an investigation would require significantly more QPU access time and lies beyond the scope of this study.

\begin{table}[h]
\centering
\caption{Calibration statistics for \texttt{ibm\_torino} machine over the months preceding and succeeding the experiments, for qubit $q_0$}
\label{calibration}
\begin{tabular}{|l|c|c|c|}
\hline
\textbf{Metric} & \textbf{Mean} & \textbf{Min} & \textbf{Max} \\
\hline
Readout error & 0.0226 & 0.0058 & 0.0783 \\
T1 ($\mu$s) & 140 & 79.6 & 216.9 \\
T2 ($\mu$s) & 103.3 & 70.0 & 155.8 \\
\hline
\end{tabular}
\end{table}

\subsection*{Limitations with machine scalability}
Our experiments focus primarily on detecting a statistically significant quantum signal in QPE histograms, rather than fully establishing polylogarithmic scaling of the expected runtime for a fixed acceptance parameter $k$ (see Appendix~\ref{polylog}). For each small modulus $N$, we tune window sizes and shot counts to test whether the mass in acceptance regions exceeds a uniform baseline. This is sufficient to detect coherence in the presence of noise but does not yet constitute a full scalability study.

A more rigorous assessment of scalability would have to go beyond signal detection and quantify an end-to-end polynomial runtime, including the number of independent repetitions required to recover $r$ with high confidence. For small $N$, one can tune $k$ and constant factors to obtain favorable behavior, which leaves too much latitude for subjective choices. A rigorous assessment therefore calls for experiments at substantially larger $N$ (spanning orders of magnitude in digit length), which is presently out of reach without fault-tolerant quantum hardware. Consequently, while our results demonstrate that a non-trivial quantum signal is still observable for modest $N$ on state-of-the-art devices, they do not yet support extrapolation to cryptographic key sizes.

\section{Trends in Quantum Computing Developments}
Investment in quantum technologies has surged in recent years. Global public and private funding has been estimated to exceed \$55 billion\footnote{https://www.qureca.com/quantum-initiatives-worldwide/}, spanning quantum computing, communication, sensing, and related areas. Many governments have launched national or regional initiatives, such as the U.S. National Quantum Initiative\footnote{https://www.quantum.gov/}, the European Quantum Flagship \footnote{https://qt.eu/} and major programs in China and other countries, aimed at securing leadership in this emerging and potentially transformative field~\cite{b10}. In parallel, private sector funding has grown rapidly, with large technology companies such as IBM, Google, Microsoft, and Amazon heavily investing in quantum computing, alongside startups such as IonQ, Quantinuum, Rigetti, PsiQuantum, Xanadu, and others, which are developing specialized quantum systems. 
Table \ref{roadmap} illustrates selected quantum computing roadmaps announced by industrial players, grouped by qubit technology. 

\begin{table*}[h]
\centering
\setlength{\tabcolsep}{1pt}
\renewcommand{\arraystretch}{1.05}
\begin{tabular}
{>{\raggedright\arraybackslash}p{3.0cm} *{7}{>{\centering\arraybackslash}p{2.0cm}}}

 & \textbf{2024} & \textbf{2025} & \textbf{2026} & \textbf{2027} & \textbf{2028} & \textbf{2029} & \textbf{2030} \\
\toprule
\textbf{Superconducting} \\
\mnum{sc}{IBM}        & \mnum{sc}{1.1k (phys)} &  &  &  &  & \mnum{sc}{200 (logical)} &  \\
\mnum{sc}{Google}     &  &  &  &  &  &  & \mnum{sc}{1M (phys)} \\
\mnum{sc}{IQM}        &  &  &  &  &  &  & \mnum{sc}{$\sim$1M (phys)} \\
\mnum{sc}{Rigetti}    &  & \mnum{sc}{36$\rightarrow$100+ (phys)} &  &  &  &  &  \\
\hline
\textbf{Trapped-ion} \\
\mnum{ti}{IonQ}       &  & \mnum{ti}{$\sim$100 (phys)} &  & \mnum{ti}{10k (phys)} & \mnum{ti}{20k (phys)} &  & \mnum{ti}{2M (phys); 40--80k (logical)} \\
\mnum{ti}{Quantinuum} &  &  &  &  &  & \mnum{ti}{100 (logical)} &  \\
\hline
\textbf{Neutral atoms} \\
\mnum{na}{PASQAL} &  & \mnum{na}{140+ (phys)} & \mnum{na}{10k (phys)} & \mnum{na}{20 (logical)} &  & \mnum{na}{100 (logical)} & \mnum{na}{200 (logical)} \\
\mnum{na}{QuEra}      &  &  & \mnum{na}{10k (phys) / 100 (logical)} &  &  &  &  \\
\hline
\textbf{Photonic} \\
\mnum{ph}{PsiQuantum} &  &  &  &  &  & \mnum{ph}{$\sim$1M (phys)} &  \\
\mnum{ph}{Xanadu}     &  &  &  &  &  &  &  \\
\mnum{ph}{Quandela}   &  & \mnum{ph}{24/100 (phys)} &  & \mnum{ph}{10 (logical)} & \mnum{ph}{50 (logical)} &  &  \\
\bottomrule
\end{tabular}
\caption{Company roadmaps with numbers of physical or logical qubits.}
\label{roadmap}
\end{table*}


\paragraph*{Hardware architectures}
Superconducting-qubit efforts, led by IBM, Google and others, emphasize scaling and fault tolerance through surface-code-like QEC. IBM was the first to surpass 1,000 qubits on a single chip and continues to pursue large scale architectures with a recent focus on modularity. Google has articulated a roadmap targeting around one million physical qubits by 2029, with the explicit goal of achieving fault-tolerant quantum computation through QEC.

Trapped-ion providers such as IonQ and Quantinuum focus on saller but higher-fidelity systems. In 2024, Quantinuum introduced a 56-qubit system that uses ion shuttling across a chip to perform operations pushing system level benchmarks such as quantum volume (QV)~\cite{cross_validating_2019,baldwin_re-examining_2022} to new record values. In 2025, the company released a 98 qubit next-generation system, which reached a point where the QV benchmark becomes too expensive in terms of classical computation to be carried out.  IonQ targets 64 qubits by 2026 and 1,024 qubits by 2028 by connecting separate chips via photonic links for scalability. Their recent acquisition of smaller startup Oxford Ionics makes them the record holder in terms of fidelities with a two-qubit gate above $99.99\%$ fidelity~\cite{hughes_trapped-ion_2025} and SPAM fidelities as high as $99.9993\%$~\cite{sotirova_high-fidelity_2024}, albeit this is for an architecture involving microwave control different to their original laser-based one.

Photonic-qubit companies, notably PsiQuantum and Xanadu, pursue architectures that exploit integrated optics and telecom technologies.  PsiQuantum has proposed a path to a one-million-qubit fault-tolerant computer by 2029, based on a fusion-based architecture~\cite{bartolucci_fusion-based_2023} that combines small photonic resource states in a probabilistic manner. However, major technical milestones, including the realization of large-scale, on-chip entanglement with low loss, are still outstanding.  
Xanadu is taking a different approach by using squeezed states of light, which are more resistant to signal loss. In 2022, the company showcased a prototype device that achieved a milestone similar to Google's breakthrough in quantum computational advantage~\cite{b23}. In addition to hardware, Xanadu develops the open-source PennyLane platform to support hybrid quantum-classical algorithm design.

\paragraph*{System scale, error rates and error correction}
For context, earlier resource estimates suggest that factoring a 2048-bit RSA modulus using a relatively direct implementation of Shor’s algorithm would require on the order of 8.4 million physical qubits~\cite{b13}, assuming that they can be operated with sufficiently low error rates and long coherence times. More recent analyses, which incorporate algorithmic improvements and more efficient QEC pipelines, reduce this requirement but still call for very large, high-quality devices~\cite{b17,b27,b28}.

It is important to emphasize that qubit count alone is an incomplete measure of progress. For Shor’s algorithm and many other applications, critical parameters include gate and measurement error rates, clock speed, qubit connectivity, and the overhead of the chosen QEC code. 

The best reported error rates (all in trapped ions) for single qubit operations are currently at the level of $10^{-7}$~\cite{smith_single-qubit_2025}, with the best two-qubit gates having reached the $10^{-5}$ range~\cite{hughes_trapped-ion_2025}. These numbers, however, are \enquote{hero values}. In many multi-qubit systems, effective error rates are much closer to $10^{-4}$ for single-qubit and $10^{-2}$ for two-qubit gates once full-system effects are included. In contrast, modern digital transistors exhibit error rates as low as $10^{-23}$~\cite{b15}, highlighting the vast gap between classical and quantum reliability.

Large-scale, fault-tolerant quantum computation therefore hinges on effective QEC. In recent years, experimental demonstrations have shown that QEC can, in principle, suppress logical error rates below those of the underlying physical qubits. For instance, a demonstration in 2025~\cite{b16} reported an experiment in which an error metric is reduced from $10^{-2}$ to $10^{-7}$ using 72 physical qubits. However, that demonstration relied on post-processing rather than real-time feedback and did not correct all error channels. Real-time, active QEC remains experimentally demanding and has so far only been achieved in limited settings, e.g. in a trapped-ion experiment~\cite{b24}.

The effective clock speed of a fault-tolerant quantum computer is determined not just by physical gate times, but also by the QEC cycle time and decoding latency. Connectivity plays a critical role here: architectures with more flexible connectivity can perform QEC cycles more efficiently than those constrained to local interactions.

Resource estimates for Shor’s algorithm typically assume aggressive but plausible QEC parameters. For example, the aforementioned study~\cite{b17} estimates that factoring a 2048-bit RSA modulus in about 8 hours would require approximately 20 million physical qubits, assuming a 1~\unit{\us}  surface-code cycle time. More recent work~\cite{b27} explores improved qubit architectures and decoding approaches that might reduce both qubit counts and runtime. At the same time, experimental QEC demonstrations such as~\cite{b16} show that current decoding times can be significantly longer (70~\unit{\us}) than qubit coherence times, indicating that substantial engineering and algorithmic advances are still needed before these estimates can be realized in practice.

\section{Conclusion}
We evaluated the practical performance of Shor's algorithm on noisy intermediate-scale quantum (NISQ) hardware using a public IBM quantum device. Despite significant increases in available qubits, our experiments were only able to factor very small integers such as 15 and 21 -- the same numbers that were demonstrated on much smaller quantum systems over two decades ago. 

Although many theoretical advances have improved quantum factoring algorithms in recent years, progress in  practical quantum hardware remains comparatively limited. This indicates that concerns about Shor's algorithm posing an immediate or near-term threat to currently deployed cryptographic systems are, at present, premature.

Building quantum computing systems with on the order of a million physical qubits will require substantial  advancements over the next 10 to 15 years, combining fundamental research with large-scale engineering efforts. While many companies are exploring architectures that promise lower error rates and improved connectivity, the field still faces numerous technical hurdles. Within published roadmaps and milestones, these challenges are often not fully or openly acknowledged, making independent, evidence-based analysis crucial.  

Given the wide-ranging security implications of quantum computing, it is essential to continually monitor its development and to assess emerging risks in order to safeguard communication infrastructures and to prepare for future quantum threats.

\section*{Acknowledgments}
We gratefully acknowledge Evgueni Rousselot for his initial experiments, which were instrumental in cenabling and motivating the investigations further developed in this paper. We also would like to thank Dr. Martin Ekerå for pointing out some theoretical mistakes and taking the time to give his insights.

\bibliographystyle{unsrt}
\bibliography{references}

\begin{thebibliography}{10}

\bibitem{b1}
P.~W. Shor.
\newblock Algorithms for quantum computation: discrete logarithms and
  factoring.
\newblock In {\em Proceedings of the 35th Annual Symposium on Foundations of
  Computer Science (FOCS)}, pages 124--134, 1994.

\bibitem{b2}
P.~W. Shor.
\newblock Polynomial-time algorithms for prime factorization and discrete
  logarithms on a quantum computer.
\newblock {\em SIAM Review}, 41(2):303--332, 1999.

\bibitem{b29}
M.~Roetteler, M.~Naehrig, K.~M. Svore, and K.~Lauter.
\newblock Quantum resource estimates for computing elliptic curve discrete
  logarithms.
\newblock In T.~Takagi and T.~Peyrin, editors, {\em Advances in Cryptology --
  ASIACRYPT 2017}. Springer, 2017.

\bibitem{preskill_quantum_2018}
J.~Preskill.
\newblock Quantum computing in the {NISQ} era and beyond.
\newblock {\em Quantum}, 2:79, 2018.

\bibitem{b18}
T.~Hoefler, T.~H{\"a}ner, and M.~Troyer.
\newblock Disentangling hype from practicality: On realistically achieving
  quantum advantage.
\newblock {\em Communications of the ACM}, 66(5):82--87, 2023.

\bibitem{b3}
L.~M. Vandersypen, M.~Steffen, G.~Breyta, C.~S. Yannoni, M.~H. Sherwood, and
  I.~L. Chuang.
\newblock Experimental realization of shor's quantum factoring algorithm using
  nuclear magnetic resonance.
\newblock {\em Nature}, 414(6866):883--887, 2001.

\bibitem{b4}
C.-Y. Lu, D.~E. Browne, T.~Yang, and J.-W. Pan.
\newblock Demonstration of a compiled version of shor's quantum factoring
  algorithm using photonic qubits.
\newblock {\em Physical Review Letters}, 99(25):250504, 2007.

\bibitem{b5}
B.~P. Lanyon, T.~J. Weinhold, N.~K. Langford, M.~Barbieri, D.~F. James,
  A.~Gilchrist, and A.~G. White.
\newblock Experimental demonstration of a compiled version of shor's algorithm
  with quantum entanglement.
\newblock {\em Physical Review Letters}, 99(25):250505, 2007.

\bibitem{b6}
E.~Lucero et~al.
\newblock Computing prime factors with a josephson phase qubit quantum
  processor.
\newblock {\em Nature Physics}, 8(10):719--723, 2012.

\bibitem{b7}
E.~Martin-Lopez, A.~Laing, T.~Lawson, R.~Alvarez, X.-Q. Zhou, and J.~L.
  O'Brien.
\newblock Experimental realization of shor's quantum factoring algorithm using
  qubit recycling.
\newblock {\em Nature Photonics}, 6(11):773--776, 2012.

\bibitem{b8}
T.~Monz et~al.
\newblock Realization of a scalable shor algorithm.
\newblock {\em Science}, 351(6277):1068--1070, 2016.

\bibitem{b9}
M.~Amico, Z.~H. Saleem, and M.~Kumph.
\newblock Experimental study of shor's factoring algorithm using the ibm q
  experience.
\newblock {\em Physical Review A}, 100(1):012305, 2019.

\bibitem{b30}
M.~Sobhani, Y.~Chai, T.~Hartung, and K.~Jansen.
\newblock Variational quantum eigensolver approach to prime factorization on
  {IBM's} noisy intermediate-scale quantum computer.
\newblock {\em Physical Review A}, 111(4):042413, 2025.

\bibitem{b12}
J.~A. Smolin, G.~Smith, and A.~Vargo.
\newblock Oversimplifying quantum factoring.
\newblock {\em Nature}, 499(7457):163--165, 2013.

\bibitem{b19}
D.~Willsch, M.~Willsch, F.~Jin, H.~De Raedt, and K.~Michielsen.
\newblock Large-scale simulation of shor's quantum factoring algorithm.
\newblock {\em Mathematics}, 11(19):4222, 2023.

\bibitem{b25}
F.~Boudot, P.~Gaudry, A.~Guillevic, N.~Heninger, E.~Thom{\'e}, and
  P.~Zimmermann.
\newblock Comparing the difficulty of factorization and discrete logarithm: a
  240-digit experiment.
\newblock In {\em Advances in Cryptology -- CRYPTO 2020}, pages 62--91.
  Springer, 2020.

\bibitem{b20}
C.~Chevignard, P.-A. Fouque, and A.~Schrottenloher.
\newblock Reducing the number of qubits in quantum factoring.
\newblock {\em IACR Cryptology ePrint Archive}, 2024.

\bibitem{b21}
O.~Regev.
\newblock An efficient quantum factoring algorithm.
\newblock {\em Journal of the ACM}, 72(1):10:1--10:13, 2025.

\bibitem{b26}
R.~L. Chen.
\newblock An overview of {Regev's} quantum factoring algorithm and its recent
  developments.
\newblock {\em Applied and Computational Engineering}, 110(1):161--169, 2024.

\bibitem{b22}
S.~Ragavan and V.~Vaikuntanathan.
\newblock Optimizing space in {Regev's} factoring algorithm.
\newblock {\em IACR Cryptology ePrint Archive}, 2023:1501, 2023.

\bibitem{b17}
C.~Gidney and M.~Eker{\aa}.
\newblock How to factor 2048 bit {RSA} integers in 8 hours using 20 million
  noisy qubits.
\newblock {\em Quantum}, 5:433, 2021.

\bibitem{b27}
C.~Gidney.
\newblock How to factor 2048 bit {RSA} integers with less than a million noisy
  qubits.
\newblock {\em arXiv}, 2025.

\bibitem{b28}
{\'E}.~Gouzien and N.~Sangouard.
\newblock Factoring 2048-bit {RSA} integers in 177 days with 13,436 qubits and
  a multimode memory.
\newblock {\em Physical Review Letters}, 127(14):140503, 2021.

\bibitem{b31}
M.~Eker{\aa}.
\newblock On the success probability of quantum order finding.
\newblock {\em ACM Transactions on Quantum Computing}, 5(2):1--40, May 2024.

\bibitem{b32}
M.~Eker{\aa}.
\newblock Quantum algorithms for computing general discrete logarithms and
  orders with tradeoffs.
\newblock {\em Journal of Mathematical Cryptology}, 15(1):359--407, 2021.

\bibitem{Cuccaro2004}
S.~A. Cuccaro, T.~G. Draper, S.~A. Kutin, and D.~P. Moulton.
\newblock A new quantum ripple-carry addition circuit, 2004.
\newblock arXiv:quant-ph/0410184.

\bibitem{Oumarou2022}
O.~Oumarou.
\newblock Halving the width of {Toffoli}-based constant modular addition to
  \(n+3\) qubits.
\newblock {\em Physical Review A}, 105:052436, 2022.

\bibitem{Draper2000}
T.~G. Draper.
\newblock Addition on a quantum computer, 2000.
\newblock arXiv:quant-ph/0008033.

\bibitem{Kitaev1997}
A.~Yu. Kitaev.
\newblock Quantum computations: algorithms and error correction.
\newblock {\em Russian Mathematical Surveys}, 52(6):1191--1249, 1997.

\bibitem{Cleve1998}
R.~Cleve, A.~Ekert, C.~Macchiavello, and M.~Mosca.
\newblock Quantum algorithms revisited.
\newblock {\em Proceedings of the Royal Society of London A}, 454:339--354,
  1998.

\bibitem{QiskitRuntime}
{Qiskit Contributors}.
\newblock Qiskit runtime primitives: Sampler (v2).
\newblock \url{https://qiskit.org/documentation/}.
\newblock Qiskit/IBM Quantum documentation, accessed 2025-09-10.

\bibitem{b11}
J.~Tomcala.
\newblock On the various ways of quantum implementation of the modular
  exponentiation function for shor's factorization.
\newblock {\em International Journal of Theoretical Physics}, 63(1):14, 2024.

\bibitem{b10}
E.~Parker.
\newblock {\em Commercial and Military Applications and Timelines for Quantum
  Technology}.
\newblock RAND Corporation, 2021.

\bibitem{cross_validating_2019}
Andrew~W. Cross, Lev~S. Bishop, Sarah Sheldon, Paul~D. Nation, and Jay~M.
  Gambetta.
\newblock Validating quantum computers using randomized model circuits.
\newblock {\em Physical Review A}, 100(3):032328, 2019.

\bibitem{baldwin_re-examining_2022}
Charles~H. Baldwin, Karl Mayer, Natalie~C. Brown, Ciarán Ryan-Anderson, and
  David Hayes.
\newblock Re-examining the quantum volume test: {Ideal} distributions, compiler
  optimizations, confidence intervals, and scalable resource estimations.
\newblock {\em Quantum}, 6:707, 2022.

\bibitem{hughes_trapped-ion_2025}
A.~C. Hughes, R.~Srinivas, C.~M. Löschnauer, H.~M. Knaack, R.~Matt, C.~J.
  Ballance, M.~Malinowski, T.~P. Harty, and R.~T. Sutherland.
\newblock Trapped-ion two-qubit gates with {\textgreater}99.99\% fidelity
  without ground-state cooling, October 2025.
\newblock arXiv:2510.17286 [quant-ph].

\bibitem{sotirova_high-fidelity_2024}
A.~S. Sotirova, J.~D. Leppard, A.~Vazquez-Brennan, S.~M. Decoppet, F.~Pokorny,
  M.~Malinowski, and C.~J. Ballance.
\newblock High-fidelity heralded quantum state preparation and measurement,
  September 2024.
\newblock arXiv:2409.05805 [quant-ph].

\bibitem{bartolucci_fusion-based_2023}
Sara Bartolucci, Patrick Birchall, Hector Bombín, Hugo Cable, Chris Dawson,
  Mercedes Gimeno-Segovia, Eric Johnston, Konrad Kieling, Naomi Nickerson,
  Mihir Pant, Fernando Pastawski, Terry Rudolph, and Chris Sparrow.
\newblock Fusion-based quantum computation.
\newblock {\em Nature Communications}, 14(1):912, February 2023.
\newblock Publisher: Nature Publishing Group.

\bibitem{b23}
L.~S. Madsen et~al.
\newblock Quantum computational advantage with a programmable photonic
  processor.
\newblock {\em Nature}, 606(7912):75--81, 2022.

\bibitem{b13}
R.~Van Meter and D.~Horsman.
\newblock A blueprint for building a quantum computer.
\newblock {\em Communications of the ACM}, 56(10):84--93, 2013.

\bibitem{smith_single-qubit_2025}
M.~C. Smith, A.~D. Leu, K.~Miyanishi, M.~F. Gely, and D.~M. Lucas.
\newblock Single-{Qubit} {Gates} with {Errors} at the
  \$\{10\}{\textasciicircum}\{{\textbackslash}ensuremath\{-\}7\}\$ {Level}.
\newblock {\em Physical Review Letters}, 134(23):230601, June 2025.

\bibitem{b15}
J.~F. Ziegler and W.~A. Lanford.
\newblock Effect of cosmic rays on computer memories.
\newblock {\em Science}, 206(4420):776--788, 1979.

\bibitem{b16}
R.~Acharya et~al.
\newblock Quantum error correction below the surface code threshold.
\newblock {\em Nature}, 638(8052):920--926, 2025.

\bibitem{b24}
C.~Ryan-Anderson et~al.
\newblock Realization of real-time fault-tolerant quantum error correction.
\newblock {\em Physical Review X}, 11(4):041058, 2021.

\end{thebibliography}

\appendix
\section{Theoretical Details}\label{appendix_full}
\subsection{A randomized algorithm}\label{appendix}
The discussion in this appendix focuses on quantifying the quantum signal from measured spectra, explaining how the individual components fit together, and outlining how to certify that we are operating in a polylogarithmic-time regime.

\subsection{Per-Run Success Probability}
A single order-finding attempt succeeds if (i) the QPE spectrum places the measured outcome $y$ sufficiently close to some rational $s/r$ for continued fractions (CF) to recover the true order $r$; and (ii) given the correct $r$, the classical post-processing $\gcd\!\big(a^{r/2}\!\pm\!1,N\big)$ yields a non-trivial factor. Writing $\mathcal{A}_N$ for the CF-acceptance set and $\nu_N\in(0,1]$ for the conditional success probability of the classical step, we obtain
\begin{equation}
p_{\mathrm{succ}}(N) \;=\; \Pr(\mathcal{A}_N)\,\nu_N. \label{eq:psucc_factor}
\end{equation}

\paragraph*{Deriving $\mathcal{A}_N$.}
The CF uniqueness guarantee states that if $x$ obeys
\begin{equation}
\left|x-\frac{s}{r}\right| \;<\; \frac{1}{2r^2}, \label{eq:cf_cond}
\end{equation}
then CF correctly recovers $s/r$. In QPE we have $x=y/L$, so \eqref{eq:cf_cond} is equivalent to
\begin{equation}
\left|\frac{y}{L}-\frac{s}{r}\right| \;<\; \frac{1}{2r^2}
\;\;\Longleftrightarrow\;\;
\left|\,y-\frac{sL}{r}\,\right| \;<\; \frac{L}{2r^2}. \label{eq:y_window}
\end{equation}
Thus the strict CF acceptance half-width in integer $y$-bins is
\begin{equation}
w_0 \;=\; \frac{L}{2r^2}. \label{eq:w0}
\end{equation}
We define the acceptance set as the union of these windows around all $r$ centers:
\begin{equation}
\mathcal{A}_N \;=\; \bigcup_{s=0}^{r-1}\Big\{\,y:\;\big|\,y-\tfrac{sL}{r}\,\big|<w_0\Big\}. \label{eq:AN}
\end{equation}

\paragraph*{Classical step.}
The factor $\nu_N$ is number-theoretic: given the true $r$, we require $r$ even and $a^{r/2}\not\equiv -1\pmod N$. For semiprimes and random coprime $a$, this condition holds with constant probability bounded away from zero; a conservative universal lower bound $\nu_N\ge 1/2$ is standard, and many concrete instances have $\nu_N$ closer to $1$.

\subsection{Importance of the Quantum Signal}
Under a uniform distribution over the $L$ outcomes, the mass that falls in $\mathcal{A}_N$ equals (accepted bins)$/L$. Each window contributes $2w_0$ bins, hence
\begin{align}
\text{accepted bins} \;=\; r\,(2w_0) \;=\; r\,\frac{L}{r^2} \;=\; \frac{L}{r}, \label{eq:acc_bins}\\
b \;\stackrel{\mathrm{def}}{=}\; \Pr_{\mathrm{unif}}(\mathcal{A}_N) \;=\; \frac{1}{r}. \label{eq:baseline}
\end{align}
When certifying polylogarithmic runtime from data, we target a lower bound of the form $\Pr(\mathcal{A}_N)\ge (\log N)^{-k}$ (or end-to-end $p_{\mathrm{succ}}(N)\ge (\log N)^{-k}$) for some $k>0$. With the strict windows \eqref{eq:AN}, a perfectly flat (uniform) spectrum yields $\Pr(\mathcal{A}_N)=1/r$ by \eqref{eq:baseline}. For cryptographically relevant $N$, the order $r$ can be large, so $1/r$ may be much smaller than $(\log N)^{-k}$. Thus, a flat spectrum fails the threshold by construction—there is no way for a \enquote{flat but lucky} histogram to pass.

Consequently, substantial quantum signal is crucial. We certify it by showing that the \emph{excess mass}
\begin{equation}
\Delta \;\stackrel{\mathrm{def}}{=}\; \Pr(\mathcal{A}_N)\;-\; b \;>\; 0, \label{eq:excess}
\end{equation}
where $b$ is the uniform baseline in \eqref{eq:baseline}.

\subsection{Certifying Substantial Quantum Signal in Practice}
We test whether the observed mass in $\mathcal{A}_N$ exceeds the uniform baseline $b$.

\paragraph*{Hypotheses.}
\begin{equation}
H_0:\; p \le b \quad\text{vs.}\quad H_1:\; p > b. \label{eq:hyp}
\end{equation}

\paragraph*{Test statistic and decision.}
With $h=\text{hits}$ in $\mathcal{A}_N$ over $n=\text{shots}$, let $\widehat{p}=h/n$. The one-sided binomial $p$-value is
\begin{equation}
p_{\mathrm{val}} \;=\; \Pr\!\left[X\sim\mathrm{Bin}(n,b):\,X\ge h\right]. \label{eq:pval}
\end{equation}
We declare \textbf{PASS} at level $\alpha$ if and only if $p_{\mathrm{val}}\le \alpha$. We also report the excess mass
\begin{equation}
\widehat{\Delta} \;\stackrel{\mathrm{def}}{=}\; \widehat{p}-b. \label{eq:excess_mass}
\end{equation}

\subsection{Shor in the Ideal Regime}
In the noiseless model (with $t$ as in \eqref{eq:t_choice}), the QPE outcome distribution is a uniform mixture over $r$ peaks, each described by a squared Dirichlet/Fejér kernel centered at $s/r$. A standard bound shows that each peak places at least $4/\pi^2$ of its mass inside its CF window; averaging over the $r$ peaks yields
\begin{equation}
\Pr(\mathcal{A}_N) \;\ge\; \frac{4}{\pi^2} \;\approx\; 0.405 \quad\text{(independent of $N$).} \label{eq:ideal_mass}
\end{equation}
Combining with the number-theoretic constant $\nu_N\ge c_0>0$ gives
\begin{equation}
p_{\mathrm{succ}}(N) \;\ge\; \frac{4}{\pi^2}\,\nu_N \;\ge\; c \;>\; 0, \label{eq:psucc_constant}
\end{equation}
so $\mathbb{E}[T]=O(1)$ and the overall runtime is $O((\log N)^2)$. This formalizes the statement \enquote{ideal Shor has constant per-run success}.

\subsection{Shor in the Experimental Regime}\label{polylog}
The total expected runtime is $C_{\mathrm{run}}(N)\,\mathbb{E}[T]$. With asymptotically fast modular arithmetic,
\begin{equation}
C_{\mathrm{run}}(N) \;=\; O\!\big((\log N)^2 \log\log N\big). \label{eq:crun_exp}
\end{equation}
Targeting a polylogarithmic total runtime,
\begin{equation}
\mathbb{E}[\mathrm{time}] \;=\; C_{\mathrm{run}}(N)\,\mathbb{E}[T] \;=\; O\!\big((\log N)^{2+k}\log\log N\big), \label{eq:polylog_total}
\end{equation}
it is sufficient that certification establishes a polylogarithmic number of repetitions $\mathbb{E}[T]=O\!\big((\log N)^k\big)$.
\begin{itemize}
\item \emph{Ideal} ($k=0$): $O\!\big((\log N)^2 \log\log N\big)$.
\item \emph{Mild degradation} ($k=1$): $O\!\big((\log N)^3 \log\log N\big)$.
\item \emph{Stronger degradation} ($k=2$): $O\!\big((\log N)^4 \log\log N\big)$.
\end{itemize}
Thus $k$ is the penalty exponent multiplying the per-run arithmetic cost. Passing with smaller $k$ certifies a faster overall (polylogarithmic) runtime; failing for $k=1$ but passing for $k=2$ still certifies a polylogarithmic regime, with two extra powers of $\log N$ in wall-clock time.

\subsection{Effect of Decoherence on the QPE Output Distribution}\label{effect_decoherence}
In noisy QPE, the distribution after the inverse QFT depends critically on whether
the phase register remains in a \emph{pure} coherent state or has decohered into a
\emph{mixed} state.  In the ideal pure case, the state 
$\tfrac{1}{2^t}\sum_{x} e^{2\pi i \phi x}\,|x\rangle$ yields a set of narrow
Dirichlet peaks at $x = 2^t \phi$.  Under realistic hardware noise, however, the
controlled-$U^{2^j}$ blocks lose the fine phase bits, leaving only the lowest
Fourier harmonic of the phase. This corresponds to replacing the true phase by an
effective surviving phase $\phi_{\mathrm{eff}}$.  After the inverse QFT, a
single surviving harmonic produces a broadened envelope
\[
p(x)\;\propto\;\cos^{2}\!\left(\frac{\pi\,(x - x^\star)}{2^t}\right),
\qquad x^\star = 2^t\,\phi_{\mathrm{eff}} ,
\]
centred at $x^\star$ rather than at the ideal peak locations.
For $N=21$ (order $r=6$), decoherence suppresses all low-significance phase bits,
leaving only the most significant bit (MSB) of the binary expansion of each
eigenphase.  Among the six eigenphases 
\[
\phi \in \bigl\{\,0,\tfrac16,\tfrac13,\tfrac12,\tfrac23,\tfrac56\,\bigr\},
\]
only $\phi=\tfrac12$ has a \emph{stable binary expansion}, meaning that after
all bits except the MSB are erased, the truncated expansion remains the exact
binary representation of an eigenphase of the unitary (here $0.1000\ldots$).
All other eigenphases have binary expansions whose most significant bit (MSB) alone does not correspond
to any true eigenphase. Under decoherence they therefore contract toward the
nearest stable fixed point.  As a result, the noisy phase register is driven
toward the unique coarse-grained eigenphase 
$\phi_{\mathrm{eff}}=\tfrac12$, producing a single broadened lobe centered at
$x^\star=2^{t-1}$.

For $N=15$ (order $r=4$), the situation differs: the eigenphases
\[
\phi \in \bigl\{\,0,\tfrac14,\tfrac12,\tfrac34\,\bigr\}
\]
contain \emph{two} phases with stable binary expansions, namely
$\phi=0$ with $0.0000\ldots$ and $\phi=\tfrac12$ with $0.1000\ldots$.  
After decoherence removes all but the MSB, these two phases remain fixed points
of the coarse-graining map, because truncating their binary strings yields
another exact eigenphase.  Consequently, the noisy phase register retains weight
on both $\phi_{\mathrm{eff}}=0$ and $\phi_{\mathrm{eff}}=\tfrac12$, leading to
two surviving peaks in the histogram: one at $x^\star=0\equiv 2^t$ (right-edge
wrap-around) and one at $x^\star=2^{t-1}$.

Thus the cosine--squared lobe centered at $2^{t-1}$ is not universal: it appears
only when $\phi=\tfrac12$ is the sole stable coarse eigenphase, and the observed
peak structure directly reflects which eigenphases remain fixed points of the
decoherence-induced truncation of their binary expansions.

\end{document}